\documentclass[a4paper]{article}
\pdfoutput=1 % if your are submitting a pdflatex (i.e. if you have
\usepackage{jcappub} % for details on the use of the package, please
\usepackage{longtable}
\usepackage{epsfig}
\usepackage[section]{placeins}
\usepackage{float}
\usepackage{color}

\usepackage{lineno} % line number

\usepackage[load-configurations=astronomy]{siunitx}
\DeclareSIUnit\parsec{pc}
\DeclareSIUnit\hubble{\ensuremath{h}}
\newcommand\ion[2]{#1$\;${\scshape{#2}}}%                       % ion, i.e., CII = \ion{C}{ii}
\newcommand{\sun}{\ensuremath{\odot}}%

\title{\bf  Lyman-alpha Forests cool Warm Dark Matter}

\author[a]{Julien Baur,}
\author[a]{Nathalie Palanque-Delabrouille,}
\author[a]{Christophe Y\`eche,}
\author[a]{Christophe Magneville,}
\author[b,c]{Matteo Viel}

\emailAdd{julien.baur@cea.fr}

\affiliation[a]{CEA, Centre de Saclay, IRFU/SPP,  F-91191 Gif-sur-Yvette, France}
\affiliation[b]{INAF, Osservatorio Astronomico di Trieste, Via G. B. Tiepolo 11, 34131 Trieste, Italy}
\affiliation[c]{INFN/National Institute for Nuclear Physics, Via Valerio 2, I-34127 Trieste, Italy}

\date{Received xx; accepted xx}

\abstract{ 

The free-streaming of keV-scale particles impacts structure growth on scales that are probed by the Lyman-alpha forest of distant quasars. Using an unprecedentedly large sample of medium-resolution QSO spectra from the ninth data release of SDSS, along with a state-of-the-art set of hydrodynamical simulations to model the Lyman-alpha forest in the  non-linear regime, we issue one of the tightest bounds to date, from Ly-$\alpha$ data alone, on pure dark matter particles : $m_X > 4.09 \: \rm{keV}$ (95\% CL) for early decoupled thermal relics such as a hypothetical gravitino, and  correspondingly  $m_s > 24.4 \: \rm{keV}$ (95\% CL) for a non-resonantly produced right-handed neutrino. This limit depends on the value on $n_s$, and  Planck measures a higher value of $n_s$ than SDSS-III/BOSS. Our bounds thus change slightly when Ly-$\alpha$ data are combined with CMB data from Planck 2016. The limits shift to $m_X > 2.96 \: \rm{keV}$ (95\% CL) and $m_s > 16.0 \: \rm{keV}$ (95\% CL). Thanks to SDSS-III data featuring smaller uncertainties and covering a larger redshift range than SDSS-I data, our bounds confirm the most stringent results established by previous works and are further at odds with a purely non-resonantly produced sterile neutrino as dark matter. 

}

\begin{document}
\maketitle
\flushbottom
%\linenumbers

\section{Introduction}
\label{sec:intro}
Dark Matter was introduced back in the 1930's to account for discrepancies between the apparent mass in galaxies or clusters and the necessary mass for these systems to be self-gravitating in the first place. It is broadly defined as a form of matter that has mass but does not radiate electromagnetic waves in a detectable way. Large scale structures (LSS) and the cosmic microwave background (CMB) temperature anisotropies are consistent with a total matter density of $\sim$30 \% the critical density of the Universe, whereas the element abundances predicted by Big Bang nucleosynthesis (BBN) and CMB temperature anisotropies once again, are consistent with a \textit{baryonic} matter density of 4-5\% of the critical density. It thus appears with increasingly supporting evidence that about five sixths of the total matter component in the Universe are in a \textit{non-baryonic} form.\\

Neutrinos are the only particles in the standard model of particle physics to be legitimate candidates for non-baryonic dark matter, since they only interact via the weak nuclear force and have a non-zero mass as supported by the observation of oscillations between their leptonic states. However, in order to be consistent with the 1D Lyman-alpha (\textsc{Ly}-$\alpha$) forest power spectrum, the total sum of their mass eigenstates $\Sigma m_\nu$ is currently constrained to at most 0.12 eV \cite{Palanque2015b}. This confers neutrinos large velocity dispersions at the epoch of galaxy formation, making them hot dark matter (HDM) as opposed to cold dark matter (CDM) particles, since the latter  have negligible or vanishing velocity dispersions at the epoch of structure formation. In the HDM paradigm, structure formation follows a top-down scenario, wherein large filament structures form first before fragmenting into smaller systems. In the CDM bottom-up scenario, relatively small objects collapse first before aggregating into larger systems. Given the homogeneity of density contrast at the last-scattering surface ($z \sim 1050$) to within $\sim 10^{-5}$, ultra-relativistic DM particles such as standard (left-handed) neutrinos cannot clump together to form self-gravitating bodies early enough to account for the epochs at which galactic-like structures are observed. The existence of high redshift quasi-stellar objects (QSOs) directly contradicts the HDM prediction. Given their fatal flaws, apart from mixed configurations, HDM models quickly became unfeasible.\\

Cold dark matter, despite its ringing endorsement on large scales, features discordance with observations at sub-galactic scales. Several hundreds of  galactic satellites with masses in excess of $10^8 M_\sun$ are expected in halos as massive as the Milky Way, yet only a handful of candidate dwarf galaxies are observed. This illustrates how CDM predicts vastly superior numbers of low-mass halos, \textit{i.e.}, excessive power on scales $\lesssim 10 \:  h^{-1}~\rm Mpc$, compared to power on scales $\gtrsim 30 \: h^{-1}~\rm Mpc$. Furthermore, halo cores produced in numerical simulations involving CDM are more cuspy than those inferred from rotation profiles of observed galaxies. Feedback processes and other complex astrophysical phenomenologies could account for these discrepancies. Nevertheless, more straightforward compromises between CDM and HDM were introduced in the early 1980's. Mixed dark matter (MDM, or CWDM for Cold+Warm Dark Matter) for instance, is a hybrid model in which one phase of the dark matter has ultra-relativistic velocities while the other is vanishing. Standard neutrinos are, in that sense, not ruled out as a dark matter candidate, provided they are mixed with one or several CDM components. Another alternative is to introduce a particle whose velocity dispersion at the time of structure formation is non-negligeable all-the-while lesser than that of standard HDM. The effects of warm dark matter (WDM), as it is coined, has the advantage of interpolating the effects of CDM on large scales with those of HDM at sub-galactic scales, thereby being conveniently consistent with the distribution and formation of large-scale structures, while circumventing the issues encountered by CDM that we discussed above. Amongst viable WDM particle candidates are $\rm keV$ right-handed (\textit{a.k.a.} sterile) neutrinos \cite{DodelsonWidrow94, Colombi96}. Recent detections \cite{Bulbul14, Boyarsky14, Boyarsky2015} of an unidentified feature at around 3.55 keV in the stacked X-ray spectra of the Andromeda galaxy, the Perseus cluster and stacked galaxy clusters  has sparked a lively discussion. Although the significance of the result is still debated~\citep{DracoXMM}, the authors have suggested it could be attributed to the decay of a 7.1 keV dark matter particle candidate, hinting fittingly at a right-handed neutrino. Investigation into the production mechanism of this hypothetical 7.1 keV neutrino is ongoing~\citep{Abazajian2014, Merle&Schneider}.\\

When traveling, particles can interfere with the gravitational collapse of structures \cite{DodelsonWidrow94, Colombi96, Bode2000}. Beyond their free-streaming scale, structures are affected as if the particles were at rest, whereas scales below their free-streaming scale are washed out by the particles considerable velocity dispersion. This manifests in a step-like suppression in the matter power spectrum at scales above $k_{\rm{FS}} = 2 \pi / \lambda_{\rm{FS}}$ where $\lambda_{\rm{FS}}$ is the  comoving free-streaming length of the particle, as illustrated in Fig.\ref{fig:Tk_CAMB} below, where the plot is computed from outputs of  the  \textsf{CAMB}\footnote{\tt http://camb.info} 
software~\cite{Lewis2000}, and the SPH visualizations are produced by the \textsf{Splotch}\footnote{\tt http://www.mpa-garching.mpg.de/~kdolag/Splotch} software. 
Particles of a few keV have a free-streaming scale which falls below the Mpc range and within the region probed by the Lyman-alpha forests of distant high redshift quasars. Ly-$\alpha$ forest data therefore provide an ideal tool to study keV-range WDM.\\

\begin{figure}[htbp]
\begin{minipage}{\linewidth}
\begin{center}
\epsfig{figure= 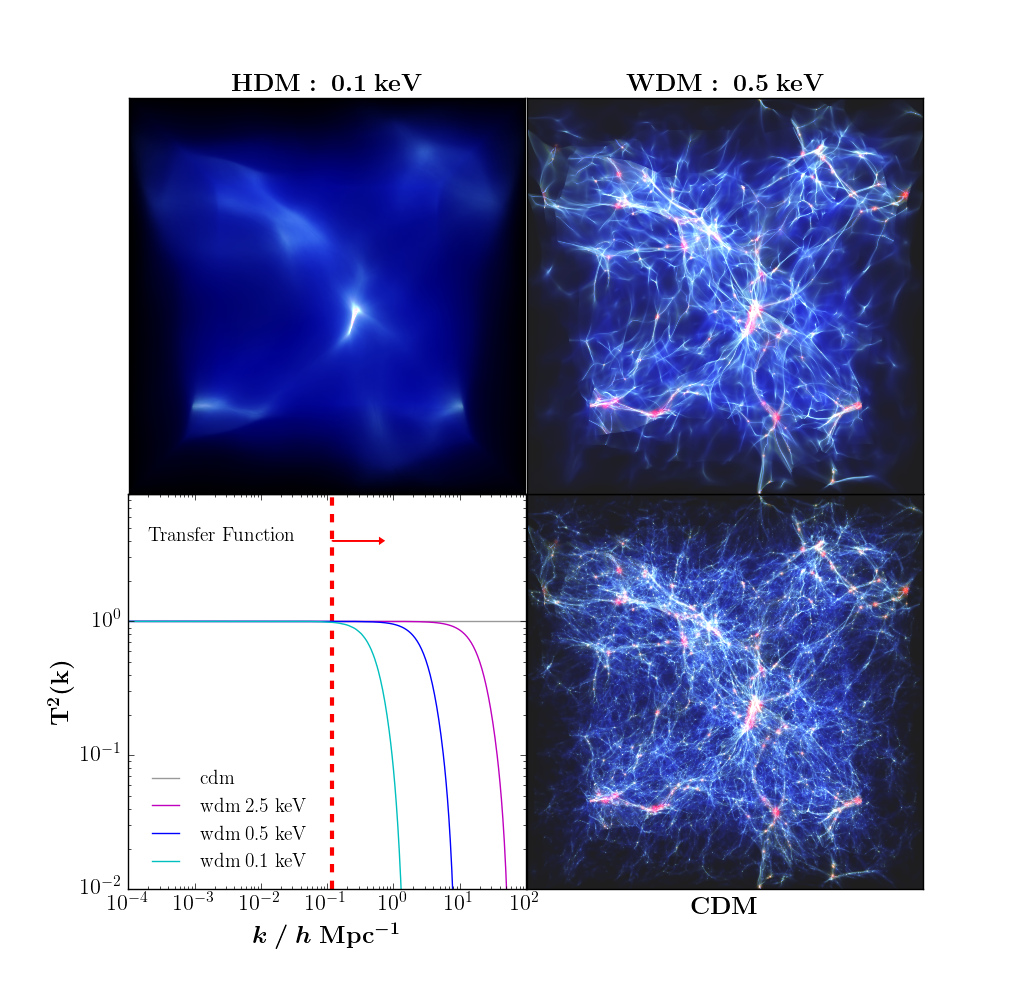, width = 16cm}
\caption{\textbf{Bottom Left:} Analytical approximation of the linear total matter transfer functions $T = \left( P_{\rm{WDM}}/P_{\rm{CDM}} \right)^{1/2}$ for 0.1 (cyan), 0.5 (blue) and 2.5 keV (magenta) dark matter particles. The region rightwards of the vertical dashed red line corresponds to the range probed by the matter power spectrum constructed from the Ly-$\alpha$ forest. The 2.5 keV case is the lightest DM mass in our grid of simulations (see Sec.~\ref{sec:simulations}), the other two being displayed for visualisation purposes. \textbf{Clockwise from Top Left:} Visual inspection of the baryon gas density and temperature (encoded in intensity and color respectively), at $z = 2.5$, for DM particle masses of 0.1 keV, 0.5 keV and for CDM (infinite mass limit). Boxes are 25 $h\rm{^{-1}Mpc}$ across in comoving coordinates and contain $768^3$ particles treated with smoothed particle hydrodynamics (baryons) and $768^3$ particles treated with N-body dynamics (DM). 
}
\label{fig:Tk_CAMB}
\end{center}
\end{minipage}
\end{figure}

Light from distant quasars excite the Lyman-$\alpha$ transition (\textit{i.e.} the neutral Hydrogen electron transitioning from the 1\textit{s} to the 2\textit{p} orbital) as it traverses matter overdensities in the IGM. The spectral redshift between the reference frames of the quasar and the absorber entails the position of the Hydrogen cloud along the observer's line-of-sight. The succession of irradiance absorption between the Ly-$\alpha$ $\lambda 1216$ and Ly-$\beta$ $\lambda 1026$ emission lines constitutes the quasar's Ly-$\alpha$ forest and a legitimate dark matter tracer at scales below $10 \: h^{-1}~\rm Mpc$. The 1D Ly-$\alpha$ flux power spectrum $P_\varphi (k) = | \tilde{\delta_\varphi} (k) |^2$, where the tilde denotes the Fourier transform and $\delta_\varphi = \varphi / \bar{\varphi} - 1$ is the transmitted flux fraction, is a well suited probe for measuring the free-streaming effect of keV-range WDM particles. To this end, the flux power spectrum established by the SDSS-III BOSS collaboration \cite{Dawson2012, Eisenstein2011, Gunn2006} from the QSO Ly-$\alpha$ forests serves to establish a lower limit on DM particle mass. Since the scales probed are well within the non-linear regime of density fluctuations, high-resolution numerical simulations are required to produce the transmitted flux power spectrum as well as to test the degeneracy of cosmological and astrophysical parameters with the mass of sterile neutrinos.\\

The outline of this article is  as follows: in Sec.\ref{sec:NuS}, we model the impact of sterile neutrinos on the matter power spectrum. Sec.\ref{sec:Lya} describes our QSO Ly-$\alpha$ forest data set from the ninth data release of SDSS quasar sample, and recaps the computation of the 1D flux power spectrum. In Sec.\ref{sec:simulations} we detail the suite of state-of-the-art hydrodynamical simulations used to construct Ly-$\alpha$ flux power spectra from a set of $\Lambda$WDM models. Sec.\ref{sec:constraints} recaps our methodology for establishing constraints on sterile neutrino masses from our simulations and data sets, along with our results which we discuss in Sec.\ref{sec:discussion}.

%%%%%%%%%%

\section{Sterile Neutrinos as WDM}
\label{sec:NuS}
While the Universe expands adiabatically, the background temperature $T_\gamma$ scales as $a^{-1}$. The background entropy density $s \propto g_{\star} T_\gamma^3$ therefore scales as $a^{-3}$, with $g_\star = \sum_{\rm{bos}} g_{\rm{bos}} + \frac{7}{8} \sum_{\rm{ferm}} g_{\rm{ferm}}$ the effective number of relativistic degrees of freedom (d.o.f.) at a given temperature. 
When the photon temperature drops below the electron mass, at $T \sim 0.5 \: \rm{MeV}$, the electrons and the positrons 
annihilate, heating the photon temperature.
Neutrinos, having decoupled prior to electron-positron annihilation ($T\sim 1 \: \rm{MeV}$), conserve the temperature $T_\nu = \alpha T_\gamma$ where $0 < \alpha < 1$. Since entropy is conserved,
evaluating the effective number of relativistic d.o.f. before ($g_\star = 11/2$) and after ($g_\star = 2$) the electron-positron annihilation yields $\alpha = (4/11)^{1/3}$. Neutrinos remain relativistic until their temperature drops below the $\rm{eV}$ scale. Their energy density $\rho_\nu$ is  given by
\begin{equation}
\label{eq:Neff}
\frac{\rho_\nu}{\rho_\gamma}= \frac{7}{8} \left( \frac{T_\nu}{T_\gamma} \right)^4 N_{\rm{eff}}\, ,
\end{equation} 
where $N_{\rm{eff}} = 3.046 + \Delta N_{\rm{eff}}$ is the effective number of neutrino species and $\Delta N_{\rm{eff}}$ its departure from the fiducial value predicted by particle physics. The slight excess in $N_{\rm{eff}}$ above 3 accounts for the portion of neutrinos still coupled to photons when electron-positron annihilation heats the background temperature. Any particle of mass $m$ and temperature $T$ coupled to photons prior to neutrino decoupling would contribute a $\Delta N_{\rm{eff}} \propto (T/T_\nu)^4$ times a suppression factor $\chi \in \left[ 0; 1 \right]$ that accounts for non-thermally produced particles. With these notations and  following the framework of \cite{Colombi96}, the velocity-space distribution function of particles making up the DM can be expressed as
\begin{equation}
\label{eq:DistributionFunction}
f (v) = \frac{\chi}{1+e^{p/T}} 
\end{equation} 
in a system of units in which $\hbar = 1$, $c=1$ and $k_b = 1$. The velocity $v$ of the DM particle is $v = \left( 1 + (m/p)^2 \right)^{-1/2}$ where $p$ is the particle's momentum. In this scope, hot (\textit{resp.} cold) DM are the asymptotic cases $m \rightarrow 0$ (\textit{resp.} $m \rightarrow \infty$). Positive masses interpolate between these two extreme cases and constitute one of 3 parameters that define the Fermi distribution function of a WDM particle, the other two being $\alpha$ and $\chi$. Integrating Eq. \ref{eq:DistributionFunction} yields
\begin{equation}
\label{eq:OTXrelation}
\frac{m^{\rm{eff}}}{m} = \chi \left( \frac{T}{T_\nu} \right)^3\, ,
\end{equation} 
where $m^{\rm{eff}} = 94.1 \: \rm{eV} \times \omega$ corresponds to the mass of the DM particle were it a standard model (left-handed) neutrino, and $\omega = \Omega_{\rm{DM}} h^2$ is the dark matter energy density in units of critical density $\rho_{c} = 3H^2/8 \pi G$. Thermal relics do not feature a suppression factor ($\chi = 1$) since they are produced in thermal processes, whereas sterile neutrinos do feature a suppressed distribution function with $T = T_\nu$. The suppression factor is a function of the active-sterile mixing angle $\vartheta$ and lepton asymmetry parameter $\mathcal{L} = \left( n_{e^{-}} - n_{e^{+}} \right) / n_{\gamma}$. We consider non-resonnantly-produced (NRP) sterile neutrinos in which no lepton asymmetry is required. In the framework of Dodelson and Widrow~\cite{DodelsonWidrow94} (`DW' hereafter), in which the sterile population is produced by oscillations with the active sector in a Seesaw mechanism in the early Universe ($T \sim 100 \: \rm{MeV}$ for keV masses), NRP neutrinos feature a quasi-thermal distribution function as showcased in Eq.~\ref{eq:DistributionFunction} (where $\chi \propto \sin^2 2 \vartheta$) despite never reaching thermal equilibrium. Thus, Eq. \ref{eq:OTXrelation} provides the distinction between the mass of a thermally-produced particle (`thermal relic' herein) $m_X$ and NRP sterile neutrino mass $m_s$ (in the DW framework):
\[ m^{\rm{eff}} = \begin{cases} \left( T/T_\nu \right)^{3} m_X  & = \left( \Delta N_{\rm{eff}} \right)^{3/4} m_X \\ \chi \: m_s & = \Delta N_{\rm{eff}} \: m_s \\ \end{cases} \]
For a fixed DM energy density $\omega$, a given value of $\Delta N_{\rm{eff}}$ determines the free-streaming cutoff scale, as displayed in Figure \ref{fig:Tk_CAMB}. The relation $\omega \times m_s^3 \propto m_X^4$ sets a correspondance between $m_X$ and $m_s$ that registers an identical matter transfer function. Figure \ref{fig:DNeff_MxMs} below illustrates this mass mapping as functions of $\Delta N_{\rm{eff}}$, given $\Omega_{\rm{DM}} = 0.26$ and $h = 0.675$. It is noteworthy that the heavier the DM particle mass, the lesser the departure from the canonical value of $3.046$ effective neutrino species, the infinite mass limit being consistent with the CDM case. Also note that despite strong tension between $N_{\rm{eff}} = 4$ and the 2015 Planck collaboration's 95\% likelihood interval \cite{Planck2015} (see also \cite{Palanque2015a}), thereby providing evidence to the non-existence of a fourth neutrino thermalized with active neutrinos, keV sterile neutrinos as pure DM are consistent with current constraints on $N_{\rm{eff}}$. 

\begin{figure}[htbp]
\begin{center}
\epsfig{figure= 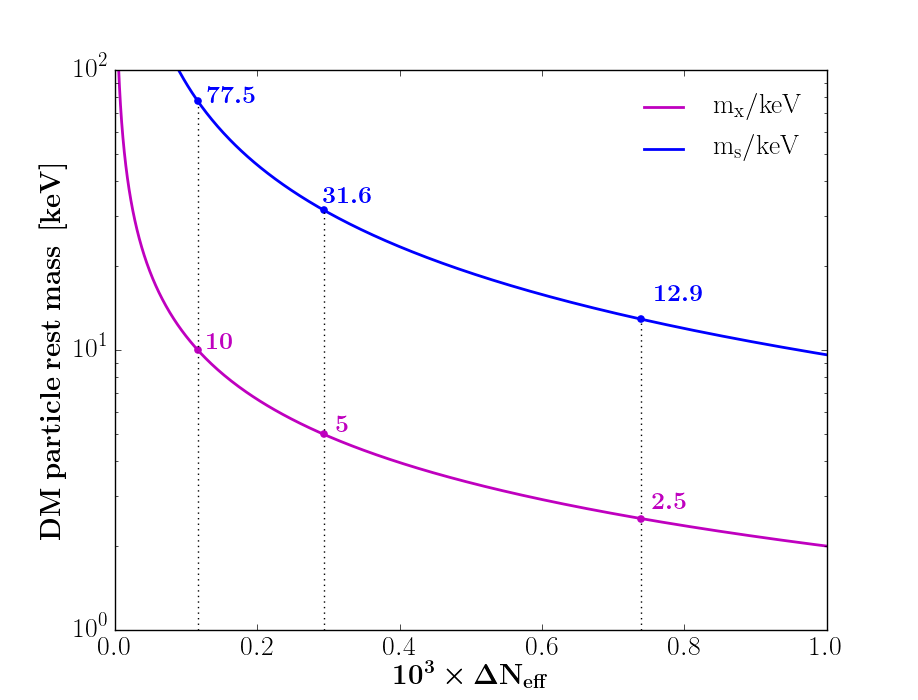, width = 12cm}
\caption{ Dark matter particle mass as a function of $\Delta N_{\rm eff}$ in the thermal relic (magenta) and DW sterile neutrino (blue) cases. The $m_X = 2.5, 5, 10 \: \rm{keV}$ values selected for our simulation grid of parameters are shown as vertical dotted lines to illustrate the one-on-one correspondance.}
\label{fig:DNeff_MxMs}
\end{center}
\end{figure}

Given the  relation between $m_s$ and $m_X$, it is sufficient to incorporate thermal relic masses $m_X$ in our grid of simulation parameters described in Sec. \ref{sec:simulations} to constrain both types of DM particles using Ly-$\alpha$ data; running an additional set of simulations for sterile neutrinos is not necessary. Given a limit on $m_X$ (or a detection!), we then  derive the corresponding range for $m_s$ using Eq.~\ref{eq:MsMxrelation} (from \cite{Abazajian2016}): \\
\begin{equation}
\label{eq:MsMxrelation}
\frac{m_s}{3.90 \rm{keV}} = \left( \frac{m_X}{\rm{keV}} \right)^{1.294} \left( \frac{0.25 \times 0.7^{2}}{\omega} \right)^{1/3}
\end{equation}

The above relationship differs from the one of \cite{VLH08a} in which $m_X$ is raised to the power $4/3$, as expected from the above discussion. The reason behind this alteration is that the DW production mechanism assumes a quasi-constant value of $g_\star$ in the range of temperature where production is most efficient. However, as the authors of \cite{Abazajian2016} point out, this approximation is not thoroughly valid and the slight fluctuations in $g_\star$ (and, more relevantly, its logarithmic derivative with respect to the scale factor $d \ln g_\star / d \ln a$) are reflected in Eq.~\ref{eq:MsMxrelation}, as the $\omega \times m_s^3 \propto m_X^4$ relation does not hold true. The $3.90$ keV numerical factor was calculated by matching the transfer functions between the thermal and non-resonant components in \cite{VLH08a} and \cite{Abazajian2006} and is accounted for in Fig.~\ref{fig:DNeff_MxMs}. \\

All of the above discussion applies to NRP sterile neutrinos in the DW framework, as their phase-space distribution function is quasi-thermal and thus exhibits thermal-like features notably in the matter power spectrum transfer function. Resonantly-produced (RP) sterile neutrinos on the other hand, such as produced in an MSW\footnote{Mikheyev-Smirnov-Wolfenstein}-like resonance introduced by Shi and Fuller~\cite{ShiFuller99}, feature a non-Fermi component in their velocity-space distribution~\citep{Abazajian2001} and therefore display a different transfer function from the one illustrated in Fig.~\ref{fig:Tk_CAMB}. Incorporating this resonant component requires running a dedicated Boltzmann code to compute the RP neutrino's phase-space distribution and transfer functions, which is beyond the scope of this work. The authors of~\citep{BLR09} have derived RP constraints from Ly-$\alpha$ forest data by approximating their transfer function at the relevant scales with a mixed Cold + Warm Dark Matter model, where the relative abundance of the cold and warm species encodes the lepton asymmetry parameter $\mathcal{L}$. We plan on following their method in a forthcoming study. Refs~\cite{Abazajian2001, ARNPS_nus} provide an extensive overview of sterile neutrinos as dark matter and their impact on cosmology given several production mechanisms.

%%%%%%%%%%

\section{Flux Power Spectrum from the Ly-$\alpha$ Forest}
\label{sec:Lya}
\begin{figure}[H]
\begin{center}
\epsfig{figure= 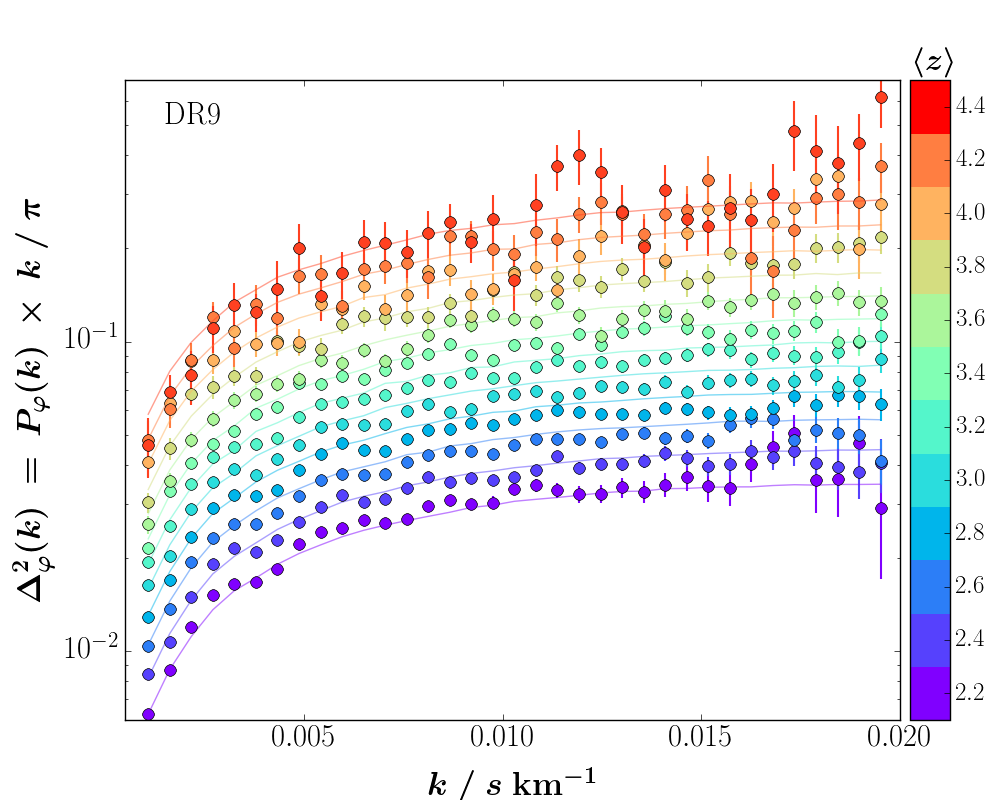, width = 12cm, height=8cm}
%figure= figure3.pdf, width = 12cm, height=8cm}
\caption{Dimensionless Ly-$\alpha$ flux power spectra $\Delta^2_{\varphi} (k) = P_{\varphi} (k) \times k/\pi$ from our selected sample in BOSS DR9. Color encodes redshift bin. Solid lines are the simulation results in each redshift bin from our benchmark model described in Sec.~\ref{sec:simulations}.
}
\label{fig:DR9fPS}
\end{center}
\end{figure}

This work is based on the one-dimensional flux power spectrum   measured using the first release of BOSS quasar data~\cite{Palanque-Delabrouille2013}.  
From a parent sample consisting of  $\sim 60,000$ SDSS-III/BOSS DR9 quasars~\cite{Ahn2012, Dawson2012, Eisenstein2011, Gunn2006, Ross2012,Smee2013}, we select the $13,821$  spectra that have  high signal-to-noise ratio, no broad absorption line features, no damped or detectable Lyman-limit systems, and an average resolution in the Ly-$\alpha$ forest of at most 
 $85 \: \rm km~s^{-1}$, where the Ly-$\alpha$ forest is defined as the region spanning  $1050 < \lambda_{RF} / \r{A} < 1180$, \textit{i.e.,} bounded by the Ly-$\alpha$ and Ly-$\beta$ emission peaks of the background quasar. The spectra in this sample are used to  measure the transmitted flux power spectrum in 12~redshift bins from $\langle z \rangle = 4.4$ to $2.2$, each bin  spanning $\Delta z = 0.2$, and in 35 equally-spaced spatial  modes  ranging from $k=10^{-3}$ to $2.10^{-2} ~\rm s~km^{-1}$ (cf. Fig.~\ref{fig:DR9fPS}). To reduce correlations between neighboring $z$-bins, we split the Ly-$\alpha$ forest of each quasar spectrum into up to three distinct redshift sectors. Each sector then has a maximum extent of $\Delta z < 0.2$.  
The fractional flux transmission is defined as $\delta_\varphi = \varphi / \bar{\varphi} - 1$, where $\bar{\varphi}$ is the mean transmitted flux fraction at the \textsc{Hi} absorber redshift, computed over  the entire sample.
The flux power spectrum is obtained from the Fourier Transform of $\delta_\varphi$, computed separately for each $z$-sector.\\

All the details  of our procedure for sample selection, calibrations, computation of the flux power spectrum and  determination of both statistical and systematic uncertainties are detailed extensively in \cite{Palanque-Delabrouille2013}. In Sec.~\ref{sec:constraints}, we  present results based on analyses covering either all 12 redshift bins (i.e., for absorber redshifts from $z=2.1$ to 4.5) or restricted to the lowest 10  bins (redshifts from $z=2.1$ to 4.1).

%%%%%%%%%%

\section{Numerical Simulations}
\label{sec:simulations}
\subsection{Simulation Pipeline}

We run a set of N-body + hydrodynamical simulations to model the flux power spectrum in the non-linear regime of density perturbations, using  \textsf{Gadget-3}, an updated version of the publicly available \textsf{Gadget-2} code\footnote{\tt http://www.mpa-garching.mpg.de/gadget/}~\cite{Springel2001,Springel2005}. 
Given a set of cosmological parameters, a user-specified box size $L$, and a number of particules per species $N^3$, the code simulates the evolution of baryonic, stellar and dark matter particles. The former undergo Smoothed Particle Hydrodynamics (SPH) treatment, a Lagrangian method of solving the hydrodynamics equations (\cite{Monaghan2005, Rosswog2009, Springel2010}), whereas the latter two are treated as collisionless populations of fixed-mass point particles. Stars are a subset of the baryon population, which the code produces whenever a particle with a temperature  less than $10^5\,{\rm K}$ reaches an overdensity with respect to the mean exceeding 1000 (as done for instance in \cite{Viel2010}). 
The code is widely used for cosmological simulations and has been thoroughly tested. 

For each simulation (\textit{i.e.,} each set of parameters), the initial matter power spectrum is generated using the \textsf{CAMB} software\footnote{\tt http://camb.info} \cite{Lewis2000} 
down to $z=30$. The power spectrum is normalized to feature a chosen fluctuation amplitude $\sigma_8$ at $z=0$ (identical for all the simulations that do not explicitly probe the impact of $\sigma_8$), before positions and velocities are generated using second-order Lagrangian perturbation theory implemented by the \textsf{2LPTic}\footnote{\tt http://cosmo.nyu.edu/roman/2LPT/} 
software. Using a splicing technique described in~\cite{McDonald2003} and first applied in~\cite{Borde2014,Palanque2015a}, we infer the flux power spectrum of an equivalent ($L=100 \:h^{-1}~ \rm Mpc$, $N=3072$) simulation from a combination of three lesser ones: a scaled-down (25, 768) to provide high resolution on small scales, a large-box low-resolution (100, 768) for large scales, and a small-box low-resolution (25, 192) which bridges the preceding two at intermediate scales. In addition to saving considerable time and resource consumption, this splicing technique loopholes around the limits of the software and packages we utilize in our pipeline. Our adaptation of the splicing technique has been successful at reproducing the `exact' spectrum for (100, 1024) and (100, 2048) configurations to within 3\% and 2\% respectively, as shown in~\cite{Palanque2015b}. 

A detailed assessment of our methodology is extensively provided in~\cite{Borde2014, Palanque2015a} and reviewed in~\cite{Palanque2015b}, along with all other simulation specifics. In what follows, we provide the reader a run-through of the major characteristics of our pipeline.\\

\subsection{Parameter Space}

Henceforth, our `best guess'  power spectrum refers to the  \textit{spliced} (100, 3072) power spectrum obtained for our central model: a flat $\Lambda$CDM Universe whose cosmological parameters are listed in the column titled `central value' of Tab.~\ref{tab:params}, in agreement with the 2013 Planck collaboration's best-fit cosmological parameters~\cite{PlanckCollaboration2013}. We produce simulations for several  values of the current expansion rate of the Universe $H_0 = 100 \: h\rm  \: km~s^{-1}~Mpc^{-1}$, the current matter energy density  $\rm \Omega_{\rm{M}}$, the spectral index of primordial density fluctuations $n_s$ and the current fluctuation amplitude of the matter power spectrum $\rm \sigma_8$. We explore two pure-$\Lambda$WDM models with $m_X = 2.5$ and 5 keV thermal relics implemented using the neutrino mass degeneracy parameters in \textsf{CAMB} to encode $\Delta N_{\rm eff} = \left( T / T_\nu \right)^4$, according to the mass-temperature relation given in Eq.~\ref{eq:OTXrelation}.  

In addition to the above cosmological parameters, we also vary parameters that describe the physics of the IGM, such as its temperature. The temperature-density relation of the IGM is measured at each redshift from the low-density regions in the simulations  and modeled with two parameters $T_0(z)$ and $\gamma(z)$ according to the relation 
\begin{equation} \label{eq:IGM}
T(\rho, z) = T_0(z) \: \left( 1 + \delta \right)^{\gamma(z) - 1}
\end{equation}
where $\delta=(\rho-\langle\rho\rangle)/\langle\rho\rangle$ is the normalized density contrast. Our central model has  logarithmic intercept $T_0^{z=3} = 14,000 \: \rm K$ and slope $\gamma^{z=3} = 1.3$, consistent with the measurements of~\cite{Becker2011}. 
Since $T_0$ and $\gamma$ are poorly constrained parameters~\cite{Garzilli2012, Lidz2010, Schaye2000}, we allow them to vary by $50 \%$ and $25 \%$ from their central values, respectively, and we run simulations for each of these sets of values. When fitting  data, however, we allow for additional freedom by modeling the  IGM temperature with three parameters describing the redshift dependence, in addition to the values of $T_0$ and $\gamma$ taken at $z=3$ (cf., Sec.~\ref{subsec:methodology} for details). 

Finally, we include in the grid two additional parameters, $A^\tau$ and $\eta^\tau$, to vary the effective optical depth of Ly-$\alpha$ absorption, although these dependences are obtained without requiring additional simulations  (see  Sec.~\ref{sec:pk} for details). The central value and range of each of the grid parameters are summarized in Tab.~\ref{tab:params}.\\

\begin{table}[htb]
\caption{List of parameters in the simulation grid, along with their range of values.}
\begin{center}
\begin{tabular}{lccl}
\textbf{Parameter} & \textbf{Central value} & \textbf{Range} & \textbf{Description} \\[2pt]
\hline \\[-10pt]
$ 1 {\rm keV} / m_X$ &  $0.0$  &  $+0.2 +0.4$ & Inverse of WDM mass when expressed in $\rm keV$ (thermal relic)\\[2pt]
$h$ &  $0.675$  &  $\pm 0.05$ & Current expansion rate in units of $100 \: \rm km~s^{-1}~Mpc^{-1}$\\[2pt]
$\rm \Omega_M$ &  $0.31$  &  $\pm 0.05$ & Current matter energy density in units of critical density\\[2pt]
$n_s$ &  $0.96$  &  $\pm 0.05$ & Scalar spectral index\\[2pt]
$\sigma_8$ &  $0.83$  &  $\pm 0.05$ & Current RMS of matter fluctuation amplitude  at $8 \:  h^{-1}~\rm Mpc$\\[2pt]
$T_0^{z=3}$ &  $\rm 14k$  &  $\pm \rm 7k$ & IGM temperature intercept at $z=3$ in K\\[2pt]
$\gamma^{z=3}$ &  $1.3$  &  $\pm 0.3$ & IGM temperature-density index at $z=3$ \\[2pt]
$A^\tau$ &  $0.0025$  &  $\pm 0.0020$ & Effective optical depth amplitude \\[2pt]
$\eta^\tau$ &  $3.7$  &  $\pm 0.4$ & Effective optical depth index \\[2pt]
\hline \\[-10pt]
\end{tabular}
\end{center}
\label{tab:params}
\end{table}

We evaluate the variations of the flux power spectrum with our set of parameters 
$$\vec{x} =  \left({\rm{keV}}/m_X, \: h, \: \rm{\Omega_M}, \: n_s, \: \sigma_8, \: T_0^{z=3}, \: \gamma^{z=3}, \: A^\tau, \: \eta^\tau \right)^T$$ 
around our central model $\vec{x}_0$ using a second-order Taylor expansion:
\begin{equation} \label{eq:Taylor}
f(\vec{x}_0 + \vec{\Delta x}) \simeq f(\vec{x}_0) + \sum_i \partial_i f(\vec{x}_0) \Delta x_i + \frac{1}{2}  \sum_{i,j} \partial^2_{ij} f (\vec{x}_0) \Delta x_i \Delta x_j \;.
\end{equation}
We run a simulation for each parameter and each value in Tab.~\ref{tab:params}, as well as  a cross-term simulation for every pair of parameters. Consequently, along with the best-guess configuration, our parameter space yields a total of 36 \textit{spliced} simulations to run, requiring over 4 Mhr CPU computing time. They are produced at the TGCC Curie machine at Bruy\`eres-le-Ch\^atel, France. All  simulations that do not require $ 1 {\rm keV} / m_X$ to be different from its central value ({\textit{e.g.,} 0) are common to this work and to the analysis of \cite{Palanque2015a, Palanque2015b}, and were not run anew.

\subsection{Constructing the Simulated Ly-$\alpha$ Power Spectrum}\label{sec:pk}

We extract 13 \textsf{Gadget} snapshots (output files) at equidistant redshifts between $z = 4.6$ and $z = 2.2$, and construct, for each snapshot, particle and line-of-sight (LOS) samples.The former serves to establish the temperature-density relation (see Eq.~\ref{eq:IGM})  of the baryon population, treated as a monoatomic gas, at the given redshift. \cite{Borde2014} provides details of the procedure and results. Consistent with standard unidimensional flux power studies, the  purpose of the latter sample is to compute the \textsc{Hi} effective optical depth $ \tau_{\rm eff}(z) = - \ln \langle \varphi (z) \rangle $ on 100,000 randomly-seeded lines-of-sight. 
At this stage, we fix the photo-ionization rate by rescaling the transmitted fluxes at each redshift in order to have the effective optical depth follow the empirical law
\begin{equation}
\label{optdepth}
 \tau_{\rm eff} = A^\tau \times \left( 1 + z \right)^{\eta^\tau}
\end{equation} 
where the amplitude  $A^\tau$ and the index $\eta^\tau $ take the desired values from Tab.~\ref{tab:params}. The central values of  $A^\tau$ and  $\eta^\tau $ are  in agreement with observations, and the allowed range encompasses observational uncertainties~\cite{Meiksin2009}. A previous study~\citep{Palanque2015b} constrained these parameters at $A^\tau = \left( 26 \pm 1 \right) \times 10^{-4}$ and $\eta^\tau = 3.734 \pm 0.015$ (68\% C.L.) using the Ly-$\alpha$ power spectrum in Sec.~\ref{sec:Lya} with a similar suite of SPH simulations. Although these uncertainty intervals are significantly narrower than our chosen steps for computing their derivative terms, we opt for a more conservative approach since the cited values apply to a different cosmological model ($\Lambda$-CDM$\nu$: cold dark matter and massive standard model neutrinos) than the present one ($\Lambda$-WDM).
The LOS-averaged transmitted flux power spectrum is then computed from the LOS sample by virtue of $P_\varphi (k) = | \tilde{\delta_\varphi} (k) |^2$, where  $\delta_\varphi$ is the transmitted flux computed using smoothed particle hydrodynamics. Note that the simulations cover a broader redshift range than the SDSS-III/BOSS DR9 data. The $z=4.6$ bin will not be used for the  analysis presented here, but it is available for future studies extending at higher redshift. 

Figure \ref{fig:TkFlux} compares the resulting flux power spectra for  $1 \:{\rm  keV} / m_X = 0.4$ and $0.2$  with that of the best-guess model at three redshifts within the Ly-$\alpha$ forest bounds probed by {BOSS}. As expected, heavier thermal-relic sterile neutrinos are more consistent with the standard cold dark matter model, which in this context would correspond to vanishing-velocity DM particles (tantamountly, of infinite mass). The power suppression caused by their free-streaming is clearly quantifiable in all three displayed redshifts, despite simulation shot noise and statistical uncertainties. Furthermore, the free-streaming cutoff is more prominent at higher redshifts, as clearly visible in the plots at the bottom of Fig.~\ref{fig:TkFlux}. This is because at high redshift, structure formation is less departed from the linear regime  and the power suppression is therefore nearer to the case of the analytical approximation illustrated in Fig.~\ref{fig:Tk_CAMB}. 
It is noteworthy to recall that the flux power spectrum in each simulation is normalized to the same  $\sigma_8$ value. Thus, the apparent excess of power at low-$k$ modes needs not be hastily interpreted as an actual accumulation of matter on large scales, but rather as an artifact of this normalization.
\begin{figure}[htbp]
\begin{center}
\epsfig{figure= 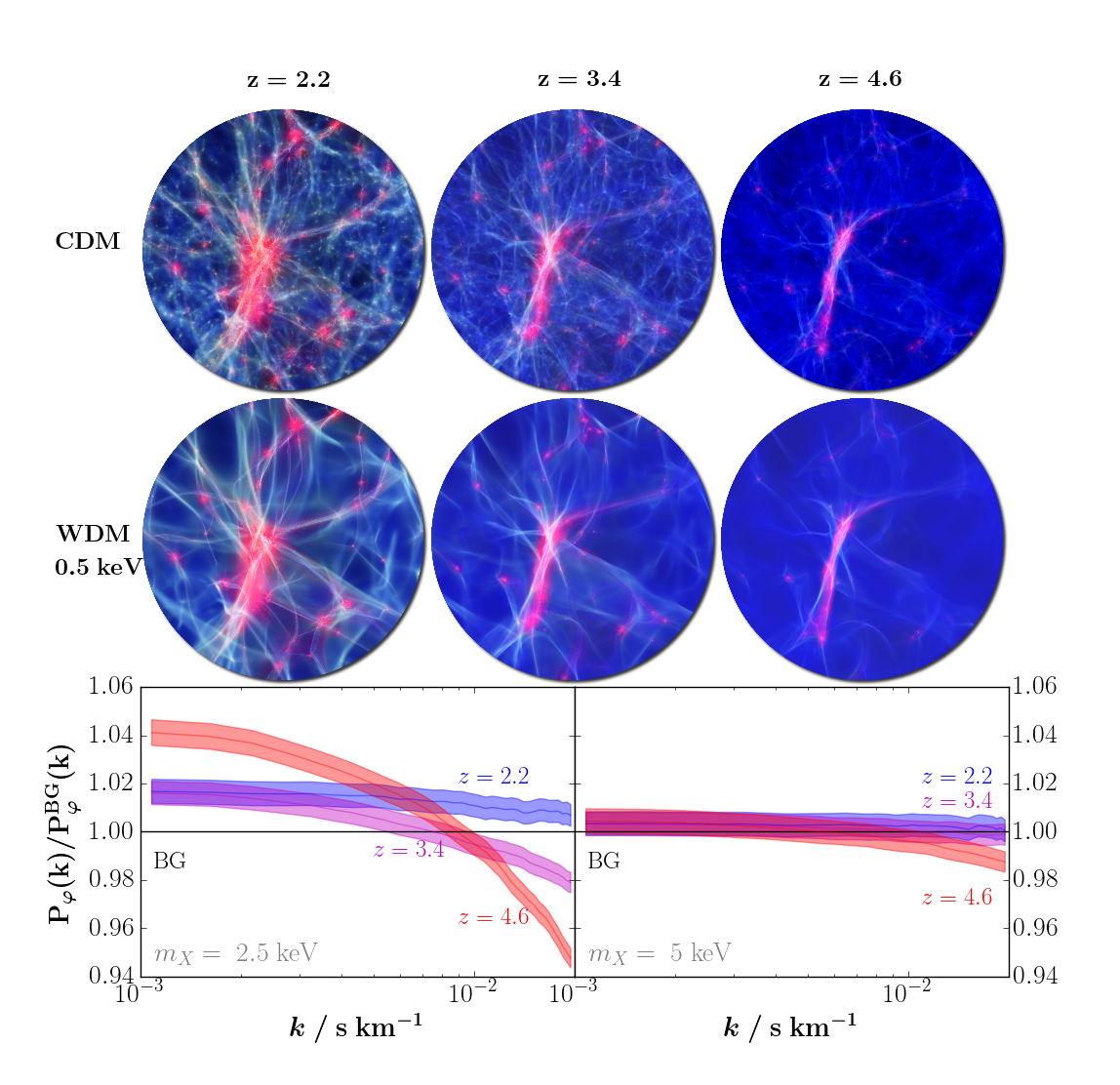, width = 16cm}
\caption{ {\bf Top \& Middle:} Visual inspection (see caption of Fig~\ref{fig:Tk_CAMB}) at $z=2.2$, 3.4 and 4.6 of the best-guess, i.e., CDM, model (top) and of a  simulation assuming a 500 eV  DM-particle mass (middle) for visualization purposes. Panels are 8  $ h^{-1}~\rm Mpc$ across. {\bf Bottom:} Ratio of the WDM to the CDM Ly-$\alpha$ transmitted flux power spectra  at redshifts $z = 2.2, 3.4$ and $4.6$, normalized to identical $\sigma_8$, for our $m_X = 2.5$~keV (left) and $5 \rm \: keV$ (right) grid values. Line thickness encodes simulation uncertainty (statistical).}
\label{fig:TkFlux}
\end{center}
\end{figure}

%%%%%%%%%%

\section{Constraints on Warm Dark Matter Mass}
\label{sec:constraints}
We follow the same approach as described in detail in \cite{Palanque2015a} and updated in \cite{Palanque2015b}, and refer to these papers for details on the method, notation, and  nuisance parameters included in the fit. We briefly summarize below the main aspects. 

\subsection{Likelihood Approach} \label{subsec:methodology}

We interpret the 1D Ly$\alpha$-flux power spectrum using  a likelihood built around three categories of parameters which are floated in the minimization procedure.  The first category describes the cosmological model in the  case of $\Lambda$WDM assuming a flat Universe (first five lines of Tab.~\ref{tab:params}). The second category models the astrophysics within the IGM:  the relationship between the gas temperature and its density (last four lines of Tab.~\ref{tab:params}), plus two amplitudes for the correlated absorption of Ly-$\alpha$ and \ion{Si}{iii}, or  Ly-$\alpha$ and \ion{Si}{ii} visible as oscillations in Fig.~\ref{fig:DR9fPS}. The redshift evolution of $T_0(z)$ and $\gamma(z)$, parameters that describe the temperature of the IGM, are modeled with power laws proportional to $[(1+z)/4]^{\rm \eta}$, where the logarithmic slope $\eta$ is unique over the whole redshift range in the case of $\gamma(z)$ but allowed to take different values above and below a $z=3$ break for $T_0(z)$. 
Lastly, in order to describe the imperfections of both our measurement of the 1D power spectrum (data) and its modeling by our suite of N-body and SPH simulations, we introduce a third category where we group all nuisance parameters that are fitted simultaneously with the parameters of interest (first two categories). This third category is implemented directly in the likelihood through well-chosen analytical  forms where free parameters give flexibility to each specific item. Theses parameters are described in Sec.~\ref{subsec:nuisance}. 

The $\chi^2$ minimization procedure (or, equivalently,  likelihood maximization) is done using the MINUIT  package~\cite{Minuit}. Our determination  of the coverage intervals of unknown  parameters is based on the `classical' confidence level method originally defined by \cite{Neyman1937}. We first determine the $\chi^2$ minimum $\chi^2_0$,  leaving  all $n$ parameters  (from the above three categories) free. To set a confidence level (CL) on any individual  parameter $\theta_i$, we then scan the variable $\theta_i$:  for each fixed value of $\theta_i$, we again minimize $\chi^2$ with the remaining $n-1$ free parameters. The $\chi^2$ difference, $\Delta \chi^2(\theta_i)$, between the new minimum and  $\chi^2_0$, allows us to compute the CL on the variable, assuming that the experimental errors are Gaussian,
\begin{equation}
{\rm CL}(\theta_i) = 1-\int_{\Delta \chi^2(\theta_i)}^{\infty}  f_{\chi^2}(t;N_{\rm{dof}}) \: dt,
\label{Eq:CL}
\end{equation}
with
 \begin{equation}
 f_{\chi^2}(t;N_{\rm{dof}})=\frac{e^{-t/2}t^{N_{\rm{dof}}/2 -  1}}{\sqrt{2^{N_{\rm{dof}}}} \Gamma(N_{\rm{dof}}/2)}
\label{Eq:chi2}
\end{equation}
where $\Gamma$ is the Gamma function and the number of degrees of freedom $N_{\rm{dof}}$
is equal to 1.
This profiling method can be easily extended to two variables. In this case, the minimizations are
performed for $n-2$ free parameters and the confidence level ${\rm CL}(\theta_i,\theta_j)$ is
derived from Eq.~\ref{Eq:CL} with $N_{\rm{dof}}=2$. Given the large number of parameters in the fit, this frequentist approach presents the advantage of being much faster than a Bayesian inference, while giving results in remarkably good agreement with the latter~\cite{Yeche2006,PlanckCollaboration2014Freq}. 

In this work, we also combine at times with the $\chi^2$ derived from Planck data. We use 
 the central values and  covariance matrices available in the official 
 Planck repositories\footnote{\tt http://wiki.cosmos.esa.int/planckpla2015/index.php/Main\_Page}  
for  the cosmological parameters  ($\sigma_8$, $n_s$, $\Omega_m$,  $H_0$, $n_{\rm run}$), where $n_{\rm run}$ is the running of the scalar index of the fluctuation amplitudes, defined as $n_{\rm run} = 2\, d n_s / d \ln k$, which appears in the first-order expansion of the initial scalar power spectrum $\mathcal{P}_S$:
\begin{equation}
\label{eq:PS}
\mathcal{P}_S = \left( \frac{k}{k_\star} \right)^{n_s - 1 + \frac{1}{2} n_{\rm run} \ln \frac{k}{k_\star}}\, ,
\end{equation} 
where $k_\star = 0.05 \: \rm{Mpc}^{-1}$ is the pivot scale of the CMB. For each parameter, we assume a Gaussian CMB likelihood with asymmetric $1\sigma$ errors that we estimate on either side of the central value from the $1\sigma$ lower and upper limits, thus accounting for asymmetric contours. We validated this strategy in \cite{Palanque2015a}, where we showed that it gave similar results to a Markov-Chain Monte-Carlo approach based on the full likelihood.

\subsection{Description of Nuisance Parameters} \label{subsec:nuisance}

The aforementioned third category of parameters is a set of 21 nuisance parameters that account for residual uncertainties or corrections related to noise in the data, spectrograph resolution, modeling of the IGM temperature-density relation, imperfections in our splicing technique, additional feedbacks to take into account in our simulations, and  redshift of reionization. Since all but the redshift of reionization are identical to \cite{Palanque2015a, Palanque2015b}, we briefly describe them in the following subsection, while we dedicate \ref{subsec:z_reio} to our treatment of the redshift of reionization.

\subsubsection{Data, Technical and Simulation Nuisance Parameters}

We model a possible redshift-dependent correction to the spectrograph resolution with the multiplicative factor 
\begin{equation}
\label{eq:Nuis1}
\mathcal{C}_{\rm{reso}} = e^{ - \left( \alpha_{\rm{reso}} + \beta_{\rm{reso}} (z-3) \right) \times k^2}
\end{equation} where $\alpha_{\rm{reso}}$ and $\beta_{\rm{reso}}$ are allowed to vary around a null value with a Gaussian constraint of $\sigma = \left( 5 \rm{km} s^{-1} \right)^2$. We also quantify the uncertainty in each 12 redshift bins of the data by multiplicative factors $\alpha^{\rm{noise}}_{\langle z \rangle}$ where $\langle z \rangle = 2.2, 2.4, ..., 4.4$. This totals 14 free parameters accounting for data uncertainty and spectrograph resolution in our likelihood.

The hydrodynamics simulations we describe in Sec.~\ref{sec:simulations} are used to compute the 1D Ly-$\alpha$ flux power spectrum from a neutral Hydrogen field. A number of astrophysical feedback processes are poorly quantitatively today and require the addition of systematics. We model the impact of feedbacks from Active Galactic Nuclei (AGN) and Supernov\ae (SN) on the Ly-$\alpha$ transmitted flux by implementing the multiplicative factors
\begin{equation}
\label{eq:Nuis2}
\begin{split}
& \mathcal{C}^{\rm{feedback}}_{\rm{AGN}} (k) = \left( \alpha_{\rm{AGN}}(z) + \beta_{\rm{AGN}}(z) \times k \right) \times \alpha^{\rm{feedback}}_{\rm{AGN}} \\
& \mathcal{C}^{\rm{feedback}}_{\rm{SN}} (k) = \left( \alpha_{\rm{SN}}(z) + \beta_{\rm{SN}}(z) \times k \right) \times \alpha^{\rm{feedback}}_{\rm{SN}}
\end{split}
\end{equation} where the $\alpha_{\rm{AGN, SN}}$ and $\beta_{\rm{AGN, SN}}$ coefficients are derived from \cite{Feedbacks}. An additional parameter is implemented to account for fluctuations in the intensity of the ionizing background, commonly referred to as UV fluctuations. Similar to \cite{UVbackground}, we implement an additive correction proportional to the transmitted flux power spectrum at  pivot point $k_{p} = 0.009 s \: \rm{km}^{-1}$, $\mathcal{C}_{\rm{UV}}$, which is $k$-independent but evolves with redshift proportionally to the power spectrum. Finally, we account for any damped Lyman-alpha (DLA) systems we might not have removed in our pipeline by introducing a $k$-dependent multiplicative correction (see \cite{McDonald2005} for justification of the analytical form)
\begin{equation}
\label{eq:Nuis3}
\mathcal{C}_{\rm DLA}(k) = \left( \frac{1}{15,000 k - 8.9} + 0.018 \right) \times 0.2 \times \alpha_{\rm{DLA}}
\end{equation} where $\alpha_{\rm{DLA}}$ is free to vary in the likelihood fit. 

We recall that the simulated flux power spectrum is obtained by a splicing technique that consists in constructing the power spectrum from a large-box simulation corrected for its low resolution, and a high-resolution simulation corrected for its small size. The residuals with respect to an exact simulation are modeled by a broken line with a pivot at $k^{\rm{min}}_{L} = 2 \pi / L$ where $L= 25 h^{-1}\rm{Mpc}$ is the  box size of the high-resolution simulation  introduced in Sec.~\ref{sec:simulations}. The relationship between $h \: \rm{Mpc}^{-1}$ and $s \: \rm{km}^{-1}$ being redshift-dependent, so is the aforementioned pivot scale, which represents the scale at which the large-box simulation correction switches from a $k$-independent to a $k$-dependent factor~(\citep{Borde2014}). Similar to \cite{Palanque2015b},  the slope below the pivot scale is fixed. The vertical offset at the pivot scale $A_{\epsilon} (k=k_p)$ is let free and we allow for a redshift-dependence in the correction slope $\eta_{\epsilon}$ beyond the pivot scale. This adds two nuisance parameters to our third category of parameters in our likelihood analysis.

\subsubsection{Reionization History}
\label{subsec:z_reio}

As in most optically-thin hydrodynamics simulations, the ionizing (UV) background causes the Hydrogen to quickly become highly ionized. This is qualitatively inaccurate since reionization processes are non-instantaneous and operate inhomogeneously in space due to density contrasts. Although hydrodynamics simulations are required to tackle this shortcoming self-consistently, our study is concerned with modeling the Ly-$\alpha$ forest at $z<5$, well after reionization processes have completed. The redshift at which the UV background onsets, however, affects the Jeans smoothing scale of the baryon gas (\citep{Gnedin&Hui98}) in a manner similar to the free streaming scale of warm dark matter particles. Altering the reionization redshift $z_{\star}$ impacts the amount of time that gas pressure has to suppress small-scale density fluctuations. It is therefore necessary to explore different thermal histories of the IGM in order to lift the degeneracy between Jeans smoothing scale and WDM free streaming scale. 
Fig.~13 of \citep{McDonald2005} shows that an increase in the redshift of reionization from $z_{\star}=7$ to 17 suppresses the Ly$\alpha$ flux power spectrum in the largest  $k$-modes present in the BOSS data ($k \sim 0.02~s \: \rm{km^{-1}}$) by about 1\% at $z=2.1$ and 4\% at $z=4.0$. Given the reduced range allowed for $z_{\star}$ by recent Planck measurements, these shifts are reduced to percent-level at most. We implement another multiplicative nuisance parameter $\mathcal{C}_\star(k)$ in our likelihood to take into account the effect of $z_{\star}$ on the IGM thermal history. This parameter is given by 
\begin{equation}
\mathcal{C}_\star(k) = \alpha_\star(z) + \beta_\star(z) k+\gamma_\star(z) k^2 \;,
\end{equation}
where $\alpha_\star$, $\beta_\star$, and $\gamma_\star$ are taken from \citep{McDonald2005} and 
interpolated with respect to the central model to our range of  redshift.\\

Since the redshift of reionization is  treated as a nuisance parameter in the fit,  we add a $z_{\star} = 9.0 \pm 1.5$ prior  to our likelihood. The central value and range of this prior are defined in order to encompass the most recent measurements of the redshift of reionization: $10.5\pm 1.1$ from WMAP9+BAO+H0~\cite{WMAP9}, $9.9\pm 1.8$ from  Planck TT temperature data at all multipoles and LFI-based polarization data at low ($\ell<30$) multipoles (PlanckTT+`lowP')~\cite{Planck2015}, $10.0\pm 1.7$ when also including the $\ell>30$ HFI polarization data  (PlanckTTEE+`lowP')~\cite{Planck2015}. The latter constraints were revised to $8.11\pm 0.93$ and $8.24\pm0.88$ respectively in the latest incarnation where the `lowP' likelihood was replaced with the `SIMlow' likelihood that includes the HFI-based polarization for $\ell\leq 20$~\cite{Planck2016PolarReio}. Values ranging from 7.8 to 8.8 are  obtained by the Planck collaboration for a given choice of CMB temperature and polarization data set, when varying the model of reionization adopted~\cite{Planck2016Reio}.

\subsection{Results} \label{subsec:results}

Combining our analysis of the Ly-$\alpha$ forest flux power spectrum with the expansion rate value of $H_0 = 67.3 \pm 1.0 \; \rm{km~s^{-1}~Mpc^{-1}}$ issued by the 2015 Planck collaboration (Ly-$\alpha$ + $\rm H_0$ herein), we obtain the most stringent lower limit on WDM mass to date, set at $m_X > 4.09 \: \rm{keV}$ for thermal relics and $m_s > 24.4 \: \rm{keV}$ for DW sterile neutrinos (95\% CL), as shown in the first part of Tab.~\ref{tab:CL95_1D}. 
The fitted values of the nuisance parameters are all well within the expected range. The IGM nuisance parameters, the corrections to our model of the splicing technique and of the spectrograph resolution are all compatible with no correction at the $1\sigma$ level. The additive corrections to the estimate of the noise power spectra range from $-9\%$ to $+19\%$ with median at $-2.5\%$ and negligible correction in the redshift bins where the noise dominates over the signal (i.e., at low redshift). The IGM temperature parameters have large error bars and are thus poorly constrained by this data set. Their values are within $1-2~\sigma$ of  typical  measurements (see, e.g. \citep{Becker2011}). Optical depth amplitude and index have consistent best-fitted values with those of the $\Lambda$-CDM$\nu$ case in~\cite{Palanque2015b} although the uncertainty ranges are larger by a factor of 2 and 4 respectively: $A^\tau = \left( 25.0 \pm 2.6 \right) \times 10^{-4}$ and $\eta^\tau = 3.728 \pm 0.074$ (68\% C.L.). \\

\begin{table}[htb]
\caption{95\% CL lower bounds on thermal relic mass $m_X$, in keV, obtained with three data configurations. When Ly-$\alpha$ is combined with other datasets, the limit is derived  with (right) or without (left) running of the spectral index. In each case, the corresponding DW sterile neutrino mass (in keV) is given in parentheses (see Eq.~\ref{eq:MsMxrelation}).}

\begin{center}
\begin{tabular}{lcc}
\hline \\[-10pt]
\textbf{Data set} & \multicolumn{2}{c}{ \textbf{Lower bound on $\; \rm{\frac{m_X}{keV}} \left( \rm{\frac{m_s}{keV}} \right)$}}\\[2pt]
\hline \\[-10pt]
Ly-$\alpha$ + $H_0$ ($z \leq 4.5$) & \multicolumn{2}{c}{4.09 (24.4)} \\[2pt]
Ly-$\alpha$ + $H_0$ ($z \leq 4.1$) & \multicolumn{2}{c}{ 2.97 (16.1)} \\[2pt]
\hline \\[-10pt]
 & no running & with running\\[2pt]

Ly-$\alpha$ + Planck {\scriptsize (TT + lowP)} & 2.96 (16.0) & 4.26 (25.7)\\[2pt]
Ly-$\alpha$ + Planck {\scriptsize (TT + lowP+ TE + EE)} + BAO & 2.93 (15.8) & 4.12 (24.6)\\[2pt]
\hline \\[-10pt]
\end{tabular}
\end{center}
\label{tab:CL95_1D}
\end{table}

The primary causes for the enhancement of our limit compared to prior studies (see Tab.~\ref{tab:studies}) are mostly two-fold. Our daughter QSO sample, for one, includes over four times as many medium-resolution spectra than previous SDSS studies, with all spectra selected for their high signal-to-noise ratio and good quality, and includes more objects in the highest redshift bins. As discussed in the end of Sec.~\ref{sec:simulations} and illustrated in Fig.~\ref{fig:TkFlux}, the damping of small-scale perturbations due to free-streaming is more prominent at higher redshifts. As such, bounds on WDM particle mass are better constrained at higher redshifts, despite observations being more challenging. Whereas our predecessors used high-resolution spectra from the HIRES, UVES and MIKE spectrographs to probe higher redshifts, our QSO sample suffices to establish a competitive constraint on $m_X$ thanks to the addition of Ly-$\alpha$ forest power spectra in the $\langle z \rangle=4.2$ and $4.4$ bins (i.e., for Ly-$\alpha$ absorbers in the redshift range between $z=4.1$ and 4.5). Dropping these two bins from our sample issues $m_X \gtrsim 3.0 \: \rm keV$, which is illustrative of their significance. Despite the relatively low number of objects in these two upper $z$-bins ($26$ and $14$ respectively out of $\sim 14,000$ in total), their contribution enhances our constraint by over 30\%. Moreover, the unprecedented resolution of our SPH numerical simulations also contributes to the competitiveness of our result. Several systematic effects have been greatly reduced: the accuracy of the splicing method and the model of its residual by a scale-dependent  feature, the quantification of the sampling variance, the model of the IGM by a broken power-law and the better accounting for the Hydrogen reionization history. The improvements on these simulation nuisance parameters are fully detailed in \cite{Palanque2015b}. Finally, we also slightly extend the range  covered by the data to smaller scales, from $k=0.018$ in SDSS-I to $0.020 ~{s\: \rm{km}^{-1}}$ in SDSS-III. 

\subsubsection{Spectral Index Running}

\begin{figure}[!]
\begin{center}
\epsfig{figure = 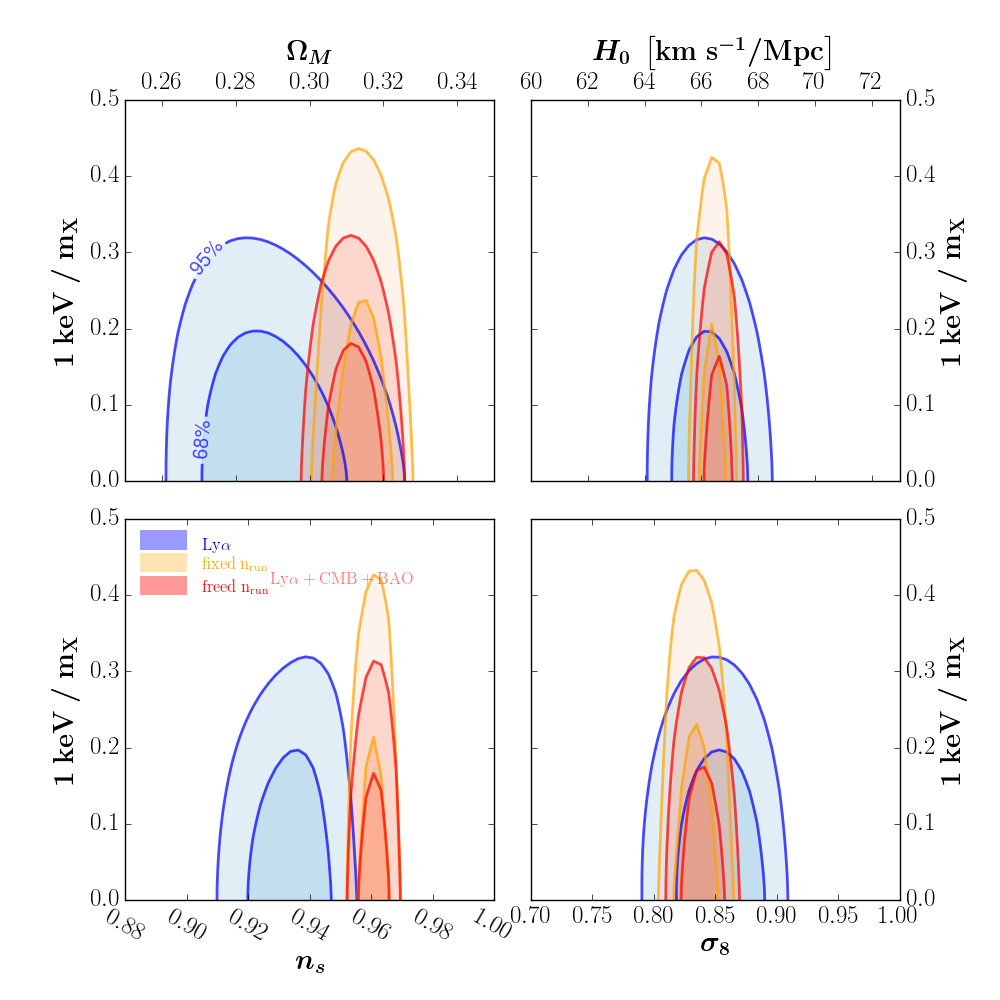, width = 16cm}
\caption{Two-dimensional probability density functions. 68\% and 95\% confidence intervals with regards to $1 \rm{keV} / m_X$ and the 4 cosmological parameters in our grid. Blue contours depict Ly-$\alpha$ forest flux power spectrum data (see Sec.~\ref{sec:Lya}) combined with the $H_0$ constraints from the Planck 2015 collaboration. Yellow and red contours are established by adding low-$\ell$ polarization, temperature and E auto and cross-correlation power spectra from Planck and measurements of the baryon acoustic oscillations scale, with the spectral index running $n_{\rm run}$ fixed to 0 (yellow) or allowed to vary (red) and fitted as a free parameter in our multidimensional analysis (see Sec.~\ref{sec:constraints}).}
\label{fig:Contour_nrun}
\end{center}
\end{figure}

In addition to Ly-$\alpha$ forests, cosmic microwave background and baryon acoustic oscillations are other formidable probes to constrain cosmological parameters. These observations cannot directly probe the small scales at which WDM plays a role, and are therefore not expected to provide a direct constraint on the mass of a WDM particle. However, they can impact our constraint on WDM mass through the correlation this parameter has with other cosmological parameters that CMB and BAO measure with a better precision than  Ly-$\alpha$ data alone can.  We thus also include, in a second step,  the Planck  temperature data at low and high multipoles (PlanckTT), the low-multipole HFI-based polarization data up to $\ell=29$ (`lowP'), and the $\ell \ge 30$ polarization  data (denoted `TE' and `EE'), as well as measurements of the BAO scale by 6dFGS \cite{6dFGS}, the main galaxy sample of SDSS \cite{SDSSmainGalaxy}, the BOSS LOW-Z sample \cite{LOWZ-CMASS} and the CMASS DR11 sample \cite{LOWZ-CMASS} (`BAO' herein). \\

Although these additional sets have contributed to establish competitive constraints on the sum of the masses of (standard) neutrinos $\Sigma m_\nu$ and the effective number of neutrino species $N_{\rm eff}$ \cite{Palanque2015a, Palanque2015b, Rossi2015}, they deteriorate our limit on WDM mass (last two rows of Tab.~\ref{tab:CL95_1D}, column labelled `no running').  This is the consequence of two factors. The first one is the  tension on the value of the spectral index  measured with  different probes: $n_s = 0.934\pm 0.009$ obtained with Ly-$\alpha$ forest and $n_s = 0.959\pm 0.004 $ with CMB. The second factor is the fact that our limit on the WDM mass is looser with increasing $n_s$. Thus the increased $n_s$ imposed by the combination with CMB is causing the loosening of our limit. \\

Ly-$\alpha$ and CMB data being relevant on  different scales, however, we can remedy the disparity on $n_s$ by allowing a non-zero running of the spectral index $n_{\rm run}$ (cf. Eq.~\ref{eq:PS}). This additional free parameter in our multidimensional analysis reconciles the different values of $n_s$ measured at small (with Ly-$\alpha$) and large (with CMB) scales. The small discrepancy on $n_s$ between Ly-$\alpha$ and CMB measurements, and the subsequent detection of $n_{\rm run}$ at  $\sim3\sigma$,  were extensively discussed in~\cite{Palanque2015b}. In the present analysis, the best-fit value of running is unchanged compared to what we measured in the context of CDM with  massive active neutrinos, which is no surprise since the detection of running is completely driven by the different values of $n_s$ measured on different scales ($n_s\sim 0.97$ at $k = 0.05~{\rm Mpc}^{-1}$ from Planck and $n_s \sim 0.94$ at $k = 0.7~{\rm Mpc}^{-1}$ from Ly-$\alpha$). We also measure no significant correlation between the value of running, which is set by the comparison of $n_s$ on large and small scales, and the value of the WDM particle mass,  set by the shape and redshift-dependence of the power spectrum on scales probed by Ly-$\alpha$ data. \\

We feature in Tab.~\ref{tab:CL95_1D} the constraints on WDM mass obtained in the `no running' and `with running' configurations, which denote the cases in which the value of the spectral index running is either taken as fixed to zero or as a free parameter, respectively. As expected, our limits on WDM mass when running is allowed to vary are similar to the limits that were derived from Ly-$\alpha$ data alone, since the effective value of $n_s$ on small scales is then determined by Ly-$\alpha$ data (and not by CMB data as in the `no running' case). We list in Tab.~\ref{tab:CL95_1D} the constraints obtained  for all three configurations (Ly-$\alpha$+$H_0$, Ly-$\alpha$+Planck with no running of $n_s$ and Ly-$\alpha$+Planck allowing for a running of $n_s$) to illustrate the impact of the value of $n_s$  on the sensitivity of our  analysis. The `with running' configuration, however, is to be considered with caution. The detection of running is driven by the different values of $n_s$ measured on large (probed by CMB) and small (probed by Ly-$\alpha$) scales. As we explained in \cite{Palanque2015b},  the determination of $n_s$ in Ly-$\alpha$ data is prone to  systematic effects in the measurement of the flux power spectrum, such as modeling of the  spectrograph resolution or contributions from SN or AGN feedbacks, UV fluctuations... The measure of $n_s$ in Ly-$\alpha$ data is  a delicate task that could still be affected by an unaccounted-for systematic. \\

Figure \ref{fig:Contour_nrun} displays our 68\% and 95\% likelihood intervals with respect to keV $/ m_X$ and our four main cosmological parameters in the `with running' (red) and `no running' (yellow) configurations. The contours for the Ly-$\alpha$ + CMB configuration are very similar to those for Ly-$\alpha$ + CMB + BAO (featured). The above discussion still holds true in the 2D case. More importantly, no significant correlation between our set of cosmological parameters and WDM mass is manifest, which conforts us in the interpretation that a small-scale power deficit in our simulated power spectrum would be due to the free-streaming of DM particles as opposed to a combined effect of $\Omega_M$, $H_0$, $\sigma_8$ and/or $n_s$. 

\subsubsection{IGM Thermal History}

\begin{figure}[!]
\begin{center}
\epsfig{figure = 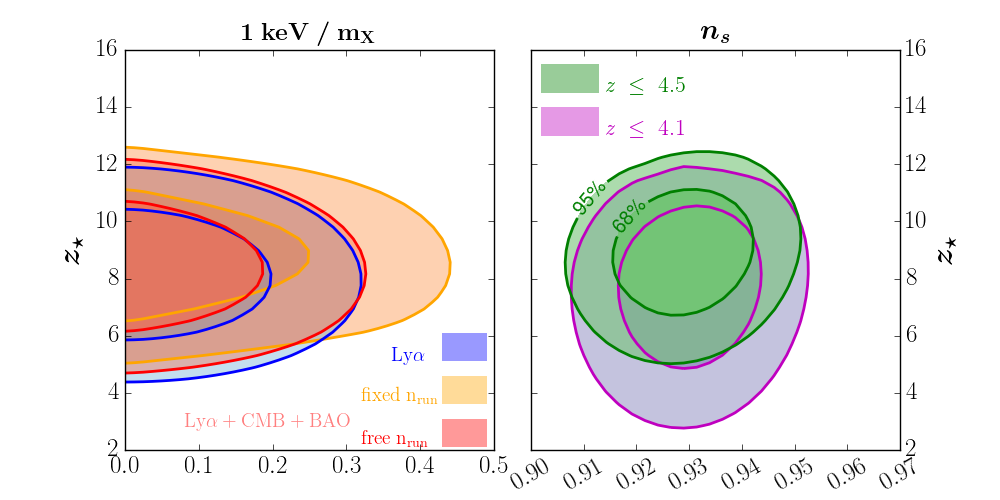, width = 16cm}
\caption{ Confidence intervals between the WDM mass (left) and the primordial spectral index (right) with the reionization redshift, given the same configurations as Fig.~\ref{fig:Contour_nrun}. The apparent lack of correlation is addressed in the text. The $z \leq 4.1$ (\textit{resp.} $4.5$) configuration corresponds to taking the first 10 (\textit{resp.} all 12) redshift bins from our data set (Ly-$\alpha$ only).}
\label{fig:Contour_Zreio}
\end{center}
\end{figure}

Our best-fit value for the redshift of reionization is $z_{\star} \simeq 8.2$ using Ly-$\alpha$ data only. The best-fit  value shifts to $z_{\star} \simeq 8.8$ and  $8.4$ in the fixed and fitted $n_{\rm{run}}$ Ly-$\alpha$+CMB configuration respectively, and to $z_{\star} \simeq 8.8$ and $ 8.5$ when BAO is also included, all consistent to within one standard deviation from the CMB constraint. 
We detect no strong correlation between $z_{\star}$ and $m_X$, as shown on the left panel of Fig.~\ref{fig:Contour_Zreio}. The global correlation coefficient of $z_\star$, defined as the correlation between that parameter and  that linear combination of all other parameters which is most strongly correlated  with it, is 60\%. This  correlation is due to the fact that several other nuisance parameters in our fit encompass --- even partially --- the effect of the IGM thermal history on the transmitted flux power spectrum, namely the splicing residual pivot offset $A_{\epsilon} (k=k_p)$ and slope $\eta_{\epsilon}$, as well as the uncertainties due to the redshift-dependence of the spectrograph resolution $\alpha_{\rm{reso}}$ and $\beta_{\rm{reso}}$. \\

Because the gas pressure has a similar effect on the power spectrum as the free streaming of WDM particles, $z_{\star}$ is expected to feature a correlation with $m_X$, as recently noted in \cite{Sherwood}. In the present situation, however, this correlation is strongly reduced for several reasons: the best-fit value for $m_X$  ventures nearby the benchmark CDM model (showing no significant departure from $1/m_X=0$), the data points with the highest statistical significance lie at low redshift where the correlation is the lowest, and the many nuisance parameters that are fitted along with the cosmology and IGM parameters also contribute to absorbing the  correlation. More generally, these nuisance parameters render $z_\star$ to have relatively small correlations with all  cosmological and astrophysical parameters. The strongest residual correlation is with the primordial spectral index $n_s$, at the $\sim 20\%$ level, which is expected since the effect of alternate values of $n_s$ on the flux power spectrum is a shift in the slope with respect to spatial scales. This slight correlation is manifest in the right panel of Fig.~\ref{fig:Contour_Zreio}, where the semi-major axis of the quasi-elliptical blue Ly-$\alpha$ contours deviates from vertical. The correlation is damped when taking CMB data into account as it probes a distinct $k$ range than Ly-$\alpha$ forest data.

\subsubsection{Warm Dark Matter or Warmer IGM ?}

\begin{figure}[!]
\begin{center}
\epsfig{figure = 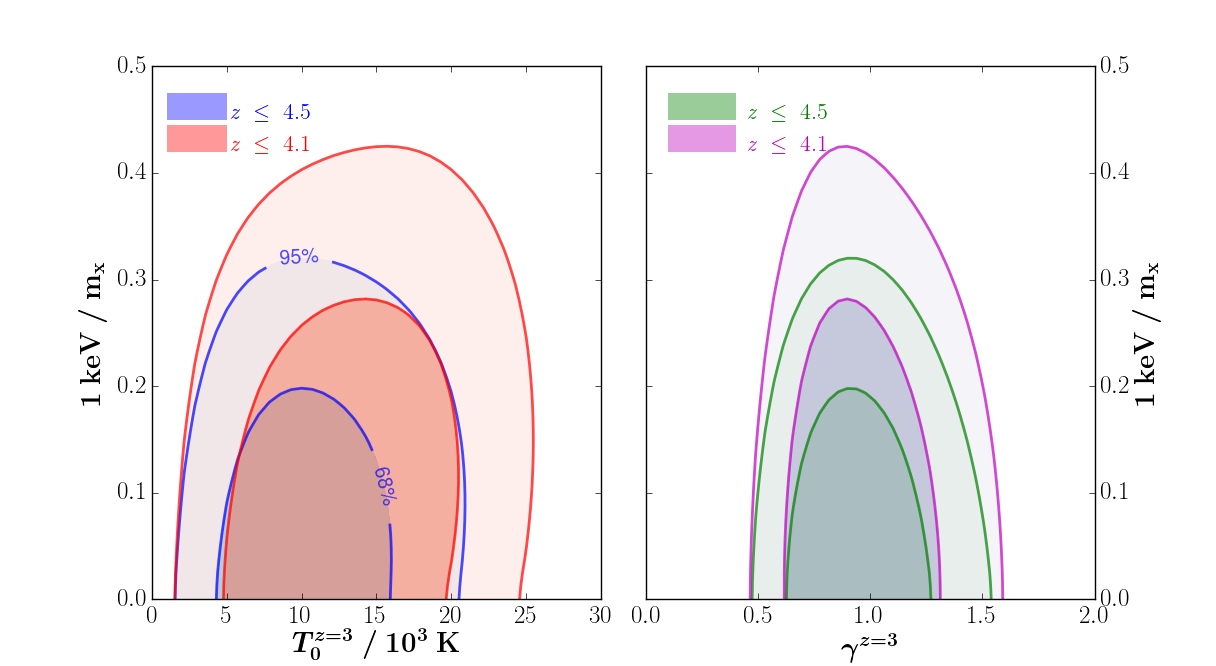, width = 16cm}
\caption{ Confidence intervals between the WDM mass and IGM temperature intercept $T_0^{z=3}$ (left) and exponent (right) $\gamma^{z=3}$ in the Ly-$\alpha$ + $H_0$ configuration. We include the confidence intervals when only taking our 10 lowest redshift bins (excluding $\langle z \rangle = 4.2$ and $4.4$, e.g., `$z \leq 4.1$' configuration) in contrast with our current extension of the analysis which features the inclusion of these two highest redshift bins (`$z \leq 4.5$' configuration).}
\label{fig:Contour_T0gamma}
\end{center}
\end{figure}

It has been recently argued in \cite{warmIGM} that the small-scale cutoff in the power spectrum can be accounted for by a warm IGM rather than a warm DM particle. The temperature-density power-law defined in Eq.~\ref{eq:IGM} is a crude first-order assumption as the power-law intercept $T_0 \: (z)$ and exponent $\gamma \: (z)$ are poorly constrained. $T_0$ may not be a monotonic function of redshift at $z \gtrsim 5$. An extended apprehension of the thermal state of the IGM and its history is crucial in carrying out investigation in the lower velocity-space $k$ segments of the Ly-$\alpha$ flux power spectrum. In our likelihood computation, we allow the IGM temperature-density relation to obey two distinct power laws, above and below a $z=3$ break.
No degeneracy between IGM temperature at $z=3$ and WDM mass is manifest, as is illustrated in Fig.~\ref{fig:Contour_T0gamma}. Implementing the $\langle z \rangle = 4.2$ and $4.4$ bins into our multidimensional analysis tightens our bounds on WDM mass and lowers $T_0 \: (z=3)$ from $\sim 14,000$ to $\sim 10,000$ Kelvins. The power-law exponent $\gamma \: (z=3)$ remains unaltered. On all accounts, the issues raised in \cite{warmIGM} do not apply to the redshift and velocity-space ranges that we probe. Our model of the IGM thermal state, although generic, is not the predominant limiting factor in the establishment of our bounds on WDM mass, as is apparent in Fig.~\ref{fig:Contour_T0gamma}. Our result is primarily limited by the sheer size and low resolution of our Ly-$\alpha$ power spectrum data sample.

%%%%%%%%%%

\section{Discussion and Conclusion}
\label{sec:discussion}
The possible existence of warm dark matter remains relevant to this day, as nuclear reactor anomalies may suggest the oscillation of active into sterile neutrinos, and recent X-ray features in galaxy clusters \cite{Bulbul14, Boyarsky14} could be interpreted as a decaying 7.1 keV pure dark matter particle. Furthermore, the most recent generations of instruments and models have ventured our progress in an era of precision cosmology. Our study is one in a long series that have attempted to use astrophysical and cosmological observations to test the plausibility of a sterile neutrino as dark matter. 
With a limit $m_X>4.09$~keV for a thermal relic or $m_s>24.4$~keV for a DW sterile neutrino, we claim the tightest constraint to date on the lower bound of WDM particle mass, using the Lyman-$\alpha$ transmitted flux power spectrum along with a set of high-resolution hydrodynamical simulations. Our work distinguishes itself from those of our predecessors mainly in the usage of a significantly larger sample of medium-resolution quasar spectra (SDSS-III) than previously (SDSS-I) and in a sharpened understanding of the systematics related to our numerical simulations, although there is undoubtedly still room for many improvements, in particular in the model of the IGM. Table \ref{tab:studies} recaps these previous lower bounds on WDM particle mass as well as the data sets and simulation resolutions to establish them. \\

The limit we obtain depends on the value of $n_s$.  Because of a  $\sim 2 \sigma$ discrepancy on $n_s$ between the value that best fits our Ly-$\alpha$ data ($n_s = 0.94\pm 0.01$) and the one issued by the Planck collaboration ($n_s = 0.966\pm0.006$), adding data from  CMB observations shifts the value of $n_s$ to the one measured by Planck and as a consequence slightly weakens our bound. The limit obtained using Ly-$\alpha$ + CMB data, $m_X>2.96$~keV for a thermal relic or $m_s>16.0$~keV for a DW sterile neutrino, is nevertheless comparable to the most stringent limits obtained recently by \cite{VBH13} or  \cite{SMT08} for instance.
As a third configuration, we reconcile the two distinct $n_s$ values by letting  the running of the spectral index vary as an additional fitting parameter. With free running, the constraints we derive on the mass of a dark matter particle are similar to the constraints we derive from Ly-$\alpha$ data alone. Since the works cited in Tab.~\ref{tab:studies} featured an implicitly null spectral index running, we issue  both our lower bound obtained with Ly-$\alpha$ data alone (with the addition of a constraint on the current expansion rate value, which Ly-$\alpha$ data is insensitive to) and the one with Ly-$\alpha$ + CMB. The subsequent addition of BAO does not affect the result. Fig.~\ref{fig:DNeff_MxMs_limits} illustrates where our bounds lie with respect to several relevant studies cited in Tab.~\ref{tab:studies}.\\

\begin{table}[htb]
\caption{Summary of past lower bounds on WDM particle mass (see text in Sec.~\ref{sec:NuS} for distinction between $m_X$ and $m_s$), along with the QSO spectrum data set used and simulation specifics. SDSS and Keck-LRIS spectra are low-resolution whereas the HIRES \cite{HIRES}, UVES and MIKE \cite{MIKE} are high-resolution. Unless explicitly stated, the simulations each contain N$^{3}$ particles for gas and for DM particles (see text in Sec.~\ref{sec:simulations}). Values for $m_s$ are not those directly quoted in the referenced works since they were computed using the (outdated) Ref.~\cite{VLH08a} relationship. We instead issue the updated value on $m_s$ using Eq.~\ref{eq:MsMxrelation} \citep{Abazajian2016} from their quoted $m_X$ value.}
\begin{center}
\begin{tabular}{lcll}
\textbf{Reference} & \textbf{95 \% limit (keV)} & \textbf{QSO spectra data set} & \textbf{Simulations} \\[2pt]
 & $m_X \: \: \: \: \: \: m_s$ &  & (L [$h^{-1}$Mpc], N) \\[2pt]
\hline \\[-10pt]
VLH05 \cite{VLH08a} & $0.55 \: \: \: \: \: \: 1.8$ & 30 HIRES + 27 UVES + 23 LRIS & (30, 200) hydro\\[2pt]
VLH06 \cite{VLH08b} & $2.0 \: \: \: \: \: \: 9.7$ &  3035 SDSS & (20, 256) hydro\\[2pt]
BLR09 \cite{BLR09} & $2.1 \: \: \: \: \: 10.4$ & 57 UVES + 3035 SDSS & (60, 400) N-body\\[2pt]
SMT06 \cite{SMT08} & $2.4 \: \: \: \: \: \: 12.2$ & 3035 SDSS & (20, 256 gas 512 DM) hydro\\[2pt]
VBH13 \cite{VBH13} & $3.3 \: \: \: \: \: \: 18.5$ & 14 HIRES + 11 MIKE & (20, 512) hydro\\[2pt]
VBH08 \cite{VBH08} & $4.0 \: \: \: \: \: \: 23.7$ & 55 HIRES + 3035 SDSS & (60, 400) + (20, 256) hydro\\[2pt]
This work & $4.1 \: \: \: \: \: \: 24.4$ & 13,821 SDSS-III & (100, 3072) hydro\\[2pt]
This work & $3.0 \: \: \: \: \: \: 16.0$ & 13,821 SDSS-III + Planck & (100, 3072) hydro\\[2pt]
\hline \\[-10pt]
\end{tabular}
\end{center}
\label{tab:studies}
\end{table}

Heavier WDM particles shift the free-streaming cutoff scale to lower scales. Therefore, furthering our constraint would require increasing the size of our quasar sample in the higher redshift bins and combining with higher-resolution spectra. Carrying out the measurement of the Ly-$\alpha$ power spectrum such as described in Sec.~\ref{sec:Lya} using an enlarged quasar sample from the twelfth data release of SDSS (eBOSS) or DESI would remedy the former. Implementing high-resolution quasar spectra from the MIKE and HIRES spectrographs alike our predecessors would accomplish the latter. Despite the relatively unbeknownst thermal state of the IGM, our modeling assumptions are generic enough to not compromise the robustness of our limit on WDM particle mass. Since the issues brought forward in \cite{warmIGM} are mostly irrelevant in the $k$ and $z$ ranges of the power spectrum we probe, our proposed future prospects are likely to further enhance the robustness of our limit.\\

\begin{figure}[htb]
\begin{center}
\epsfig{figure= 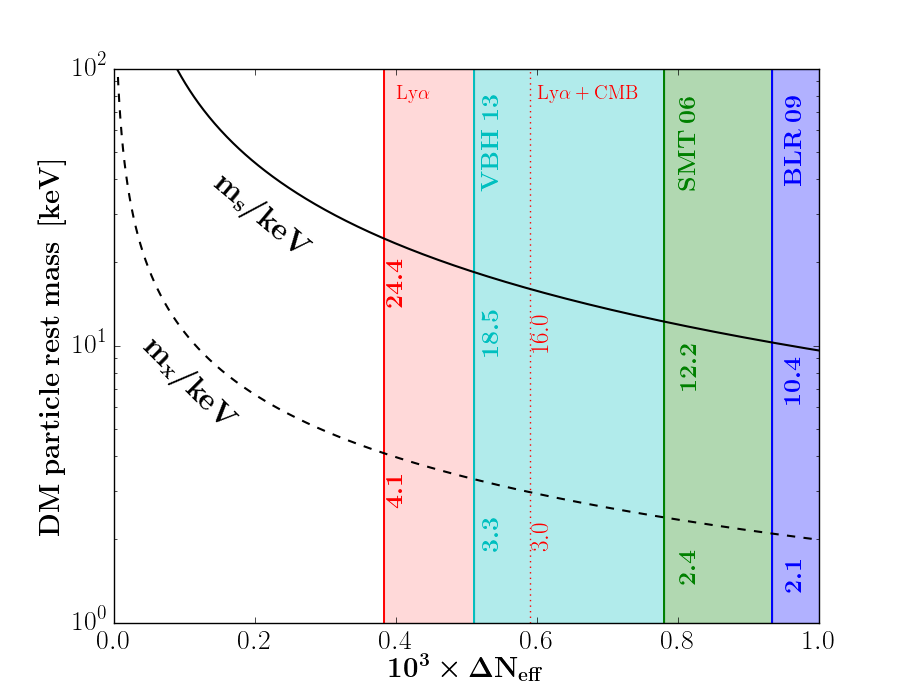, width = 12cm}
\caption{ Excluded pure-DM masses by Ly-$\alpha$ data for several past works: BLR09 \cite{BLR09, BLR09letter}, SMT06 \cite{SMT08} and VBH13 \cite{VBH13}. See Tab.~\ref{tab:studies} for a more exhaustive list. Our lower bounds, featured as a solid red line for Ly-$\alpha$ data alone and as a dashed red line for Ly-$\alpha$+CMB (with no running of $n_s$), are consistent with all previous similar studies and comparable to the most stringent limit. All cited bounds are 95\% C.L. lower limits.}
\label{fig:DNeff_MxMs_limits}
\end{center}
\end{figure}

Through this work and previous ones \cite{Palanque2015a, Palanque2015b, Borde2014, RossiNeutrino, Rossi2015}, we have demonstrated that the Ly-$\alpha$ forests of quasars are powerful measurements to probe sub-Mpc scales and that, combined with the CMB and BAO on large and intermediate scales, they have a relevant role to play in modern precision cosmology for investigating relatively minute effects such as $\Sigma m_\nu$, $N_{\rm{eff}}$, inflation slow roll parameters and the mass of WDM particles. Finally, our result is in tension with the upper bound of $m^{\rm{NRP}}_s \lesssim 4$ keV (\cite{BNRST06, BNR07}) imposed by current X-ray bounds for non-thermally produced sterile neutrinos such as the ones we have investigated in this work. We further endorse the conclusions of the works referenced in Tab.~\ref{tab:studies} in that non-resonantly produced sterile neutrinos are excluded as a pure dark matter particle. If right-handed neutrinos constitute the dark matter component, then they are either exclusively produced in resonant processes or feature both resonant and non-resonant components whose transfer function is similar to a Cold + Warm Dark Matter (CWDM) model. Bounds on resonantly-produced sterile neutrino mass from Ly-$\alpha$ forest data are established in \cite{BLR09, BLR09letter}. This work focuses on early decoupled thermal relics (such as gravitinos for instance) and neutrinos produced in a Dodelson-Widrow mechanism, and allows us to set the strongest bounds on their mass. Other models of sterile neutrinos such as the ones described above (which have been recently suggested as a plausible origin of the 3.55 keV line observed in the X-ray spectrum of galaxy clusters), will be investigated in a forthcoming study.

%%%%%%%%%%

\acknowledgments

We thank Julien Lesgourgues for his precious feedback and input, as well as Volker Springel for making \texttt{GADGET-3} available to our team. Kevork Abazajian's remarks about the neutrino to relic mass mapping were useful and appreciated. We thank James Bolton for helpful discussions on the thermal history of the IGM and its implementation in simulations.
MV is supported by the ERC-StG "cosmoIGM" and by the INDARK PD51 grant.
We acknowledge PRACE (Partnership for Advanced Computing in Europe) for awarding us access to resource curie-thin and curie-xlarge nodes based in France at TGCC, under allocation numbers 2010PA2777, 2014102371 and 2012071264.\\
\\

%%%%%%%%%%

\bibliographystyle{unsrtnat_arxiv}
\bibliography{biblio}

\end{document}